\documentclass[
  reprint,
  aps,
  prapplied,
  floatfix
  longbibliography,
  amsmath,amssymb
]{revtex4-2}

\usepackage{graphicx}
\usepackage{dcolumn}
\usepackage{bm}
\usepackage{subcaption}
\usepackage{hyperref}
\usepackage{stackengine}

\AtBeginDocument{%
  \captionsetup[figure]{justification=raggedright, singlelinecheck=false}%
  \captionsetup[subfigure]{justification=raggedright, singlelinecheck=false}%
}

\begin{document}
\preprint{APS/123-QED}
\title{A cm-wave quantum noise limited resonant superconducting parametric amplifier}


\author{V. Gilles}
\email{valerio.gilles@manchester.ac.uk}
\affiliation{Department of Physics and Astronomy, The University of Manchester, Manchester, United Kingdom}

\author{T. Sweetnam}
\altaffiliation{Present address: Oxford Quantum Circuits, Reading, United Kingdom}
\affiliation{Department of Physics and Astronomy, The University of Manchester, Manchester, United Kingdom}

\author{B. Mohammadian}
\altaffiliation{Present address: University of Oxford, Oxford, United Kingdom}
\affiliation{Department of Physics and Astronomy, The University of Manchester, Manchester, United Kingdom}

\author{M. A. McCulloch}
\affiliation{Department of Physics and Astronomy, The University of Manchester, Manchester, United Kingdom}

\author{L. Piccirillo}
\affiliation{Department of Physics and Astronomy, The University of Manchester, Manchester, United Kingdom}

\date{\today}


\begin{abstract}
Superconducting Parametric Amplifiers (SPAs) have seen great interest in recent years due to their high gain and quantum limited noise performance. Among these amplifiers, resonant SPAs have been widely developed for experiments where ultra low-noise narrow-band amplification is of interest, such as the search for Axion dark matter in particle physics and the detection of spectroscopic lines in astrophysics, while also finding applications in quantum computing.\\
This work presents an amplifier based on a Complementary Split Ring Resonator (CSRR), patterned on a NbTi coated sapphire substrate embedded within a waveguide, designed to work at a set of four narrow frequency bands throughout K band (18-27 GHz) using the kinetic inductance of the superconducting film. The S-parameters measured at 400 mK, using a sorption cooler, show the four resonances between 23.3 and 26.3 GHz at 1 GHz spacing, with a maximum transmission on resonance of -1 dB. Four-wave mixing has been observed with each resonance, and a maximum signal gain of 30 dB has been measured, corresponding to 29 dB of insertion gain. The noise performance of the amplifier has been measured, showing an added noise of 1.2 half quanta at 400 mK. These results are relevant to high-frequency Axion dark matter experiments and help motivate the exploration of higher frequencies in quantum technologies.
\end{abstract}

\maketitle

\section{Introduction}\label{Introduction}
Superconducting Parametric Amplifiers (SPAs) have been subject of extensive study in the past two decades due to their high gain and potential quantum limited noise performance of the amplifying element. These characteristics make them desirable as the first amplifier in a readout chain for a variety of applications: in physics for haloscopes \cite{DiVora2023,Bartram2023}, astrophysics for spectroscopy \cite{Westig2018} and quantum computing \cite{Aumentado2020,Gambetta2017}. The quantum limited performance of SPAs has been shown in various studies in the past decade \cite{Eom2012,Malnou2020,Castellanos-Beltran2008,Frasca_2024} but evidence of this at frequencies above 20 GHz is still lacking.\\
The two main types of SPAs are Travelling Wave Parametric Amplifiers (TWPAs) and non-linear resonators. TWPAs are non-linear transmission lines that, if long enough and implementing the correct phase-matching, can provide high wide-band gain \cite{Eom2012,Malnou2020}. Resonators can exhibit improved interaction length for compact structures and can provide narrow-band gain dictated by their quality factor \cite{Castellanos-Beltran2008,Frasca_2024}. 
The non-linearity used in these devices can be either the kinetic inductance of superconducting thin films \cite{Vissers2016,Shu2021,Eom2012} (like NbTi, NbTiN and NbN) or the non-linear inductance of Josephson junctions \cite{Planat2020,Macklin2015,Roy2015}.\\
These amplifiers can be operated both in four or three-wave mixing regimes. In four-wave mixing \cite{Anferov_2020} two pump photons interact with a signal photon to form an idler photon, with energy being transferred from the pump to the signal and idler. In the degenerate case, where the two pump photons are at the same frequency $f_p$ and the signal is at frequency $f_s$, the strongest idler is generated at the frequency: $f_i=2f_p-f_s=f_p+\Delta f$. In three-wave mixing, the pump photon gives energy to the signal and the idler, which forms at the frequency: $f_i=f_p-f_s$. Three-wave mixing has the advantage of placing the pump tone at approximately double the signal frequency, enabling the pump to be filtered out more easily than in the four-wave mixing case but often requires a DC bias to boost this interaction and achieve gain \cite{Malnou2020,Frasca_2024}.\\
For thin films in the four-wave mixing regime the kinetic inductance exhibits a quadratic dependence with the current at the lowest order \cite{Eom2012}:
\begin{equation}
L_k(I)= L_k(0) \left[ 1+(I/I_*)^2 + ...\right]
\label{Equation1}
\end{equation}
where the zero current kinetic inductance is: $L_k(0)=\hbar R_n/\pi \Delta$, with $R_n$ being the normal state resistance of the film, $\Delta$ is the superconducting gap parameter in Mattis-Bardeen theory \cite{Mattis-Bardeen1958} and $I_*$ is proportional to the critical current of the film $I_c$ ($I \ll I_*$).
The non-linearity increases with the normal state resistance of the film: $R_n=\rho_n l/A$, where $A=w \cdot t$ is the surface area of the film, $t$ is the thickness of the film, $w$ is the width, $l$ its length and $\rho_n$ is the normal state resistivity (measured in $\mathrm{\Omega \cdot m}$). These parameters can be fine tuned in order to optimise the non-linearity and achieve higher gain.\\
This work will focus on a resonator amplifier working in four-wave mixing based on a Complementary Split Ring Resonator (CSRR) structure patterned on a NbTi coated sapphire chip embedded in a waveguide sample holder\cite{Sweetnam2024}. The design of the amplifier has the advantage that the pump and signal tones can be easily coupled to the amplifying component. This allows for the possibility of patterning multiple resonating structures on the same chip, obtaining a comb-like amplification spectrum with the size of each resonator becoming the main limiting factor on the number of amplification bands.\\
Interest in operating SPAs at higher frequencies comes from Axion dark matter detection (20-50 GHz) and quantum technologies \cite{Hao2026}, where higher operating frequencies allow for the use of quantum effects at higher physical temperatures. If proven practically feasible, this could have a large impact especially in quantum computing, where most superconducting circuits have been required to use expensive and cumbersome dilution systems.

\section{Methods}\label{Methods}

\subsection{Design and Fabrication}
\begin{figure}[htbp]
  \centering
  \begin{subfigure}[b]{0.45\linewidth}
    \stackinset{l}{3pt}{t}{-14pt}{\bfseries a}%
      {\includegraphics[width=\linewidth]{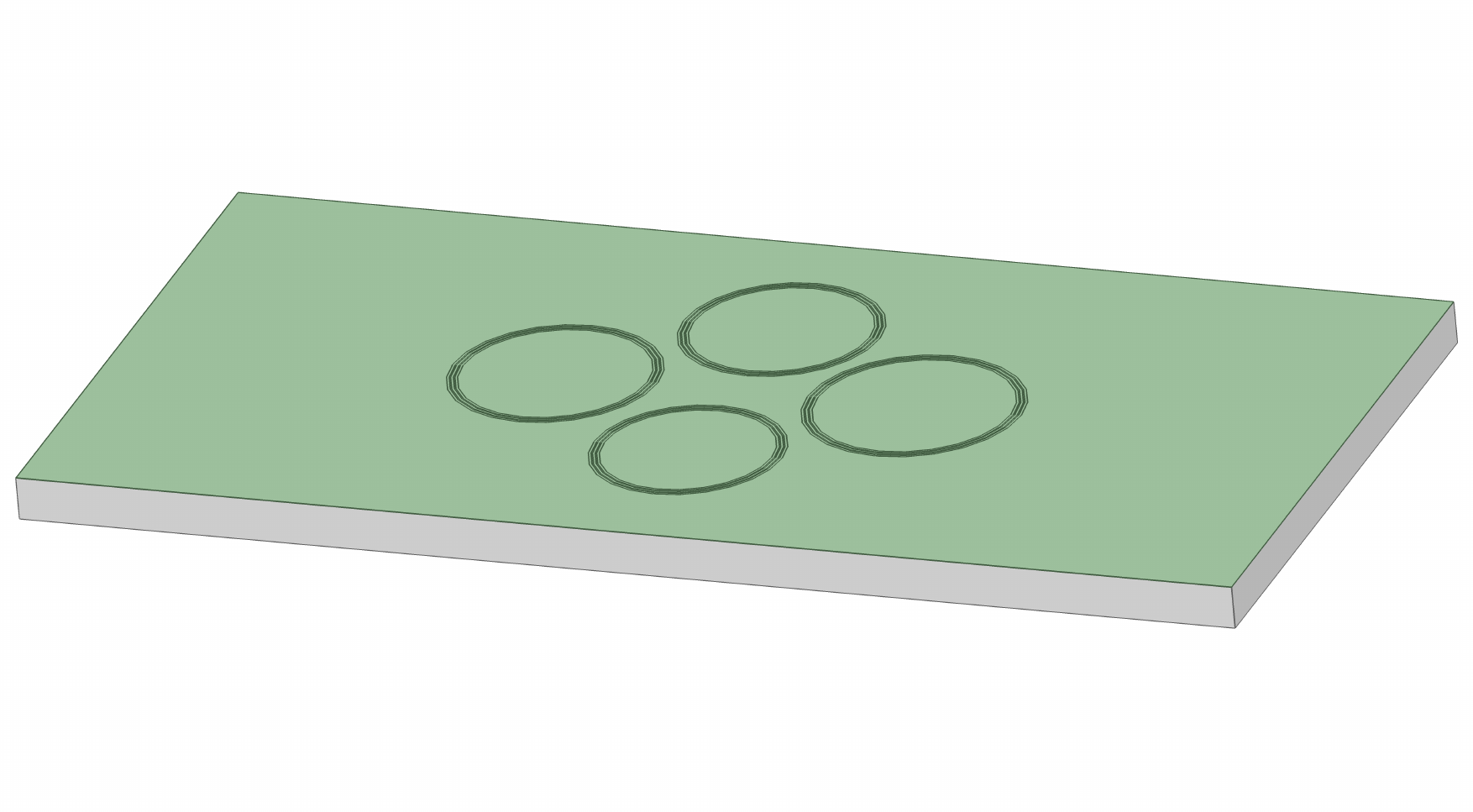}}
       \phantomsubcaption
    \label{Figure1a}
  \end{subfigure}
  \hspace{2em} 
  \begin{subfigure}[b]{0.33\linewidth}
    \stackinset{l}{3pt}{t}{4pt}{\bfseries b}%
      {\includegraphics[width=\linewidth]{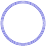}}
       \phantomsubcaption
    \label{Figure1b}
  \end{subfigure}
  \caption{(a) Chip made of a 9x5 mm sapphire substrate 250 \textrm{$\mu$}m thick coated with 42 nm of NbTi and patterned with the desired structure. (b) Design of the resonator: the white region is the NbTi, while the blue one is etched.}
  \label{Figure1}
\end{figure}

\begin{figure}[htbp]
\centering
    \begin{subfigure}[b]{0.4\linewidth}
        \centering
  \stackinset{l}{6pt}{t}{10pt}{\bfseries a}%
               {\includegraphics[width=\linewidth]{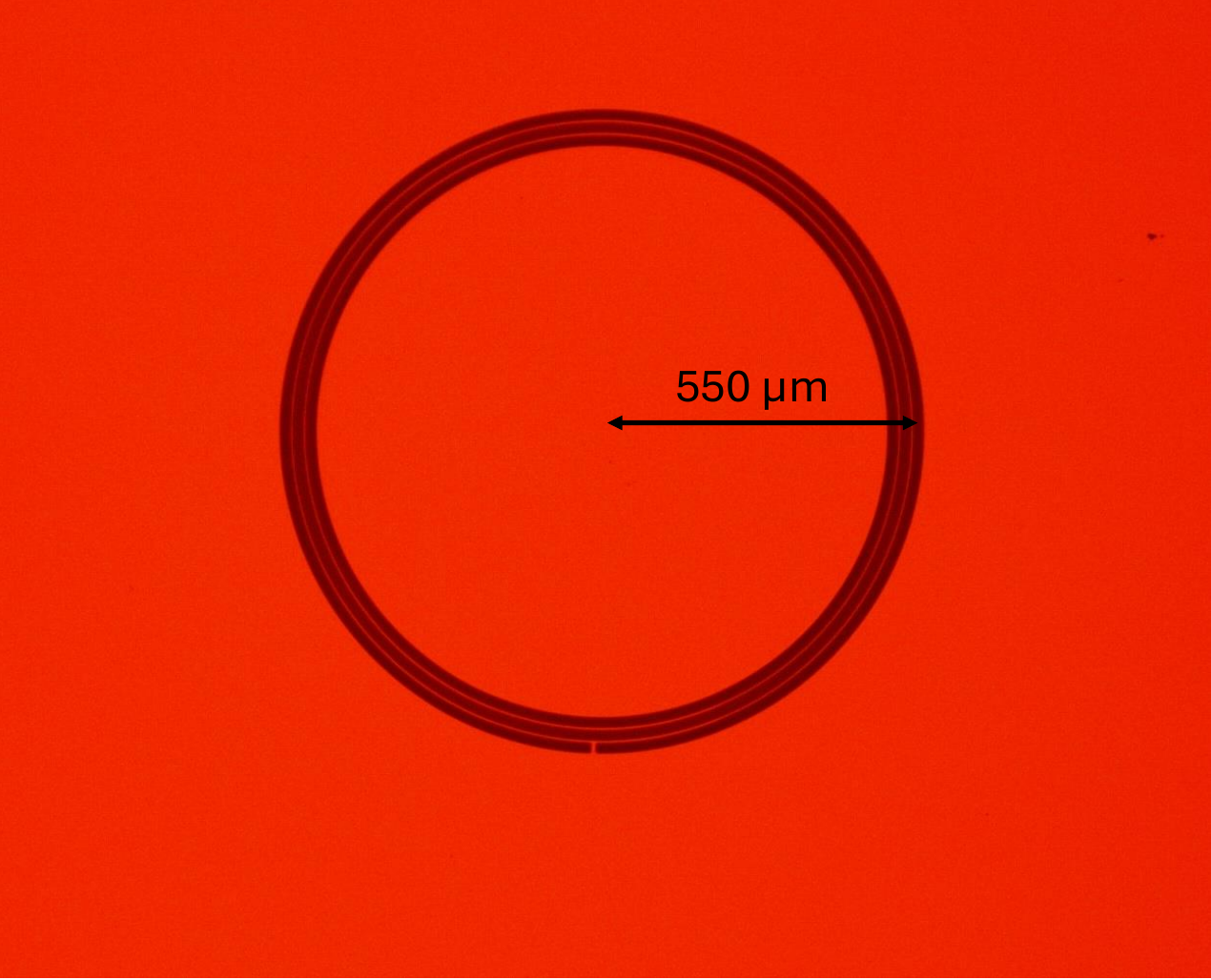}}
                \phantomsubcaption
        \label{Figure2a}
    \end{subfigure}
      \hspace{2em} 
    \begin{subfigure}[b]{0.435\linewidth}
  \stackinset{l}{6pt}{t}{9pt}{\bfseries b}%
               {\includegraphics[width=\linewidth]{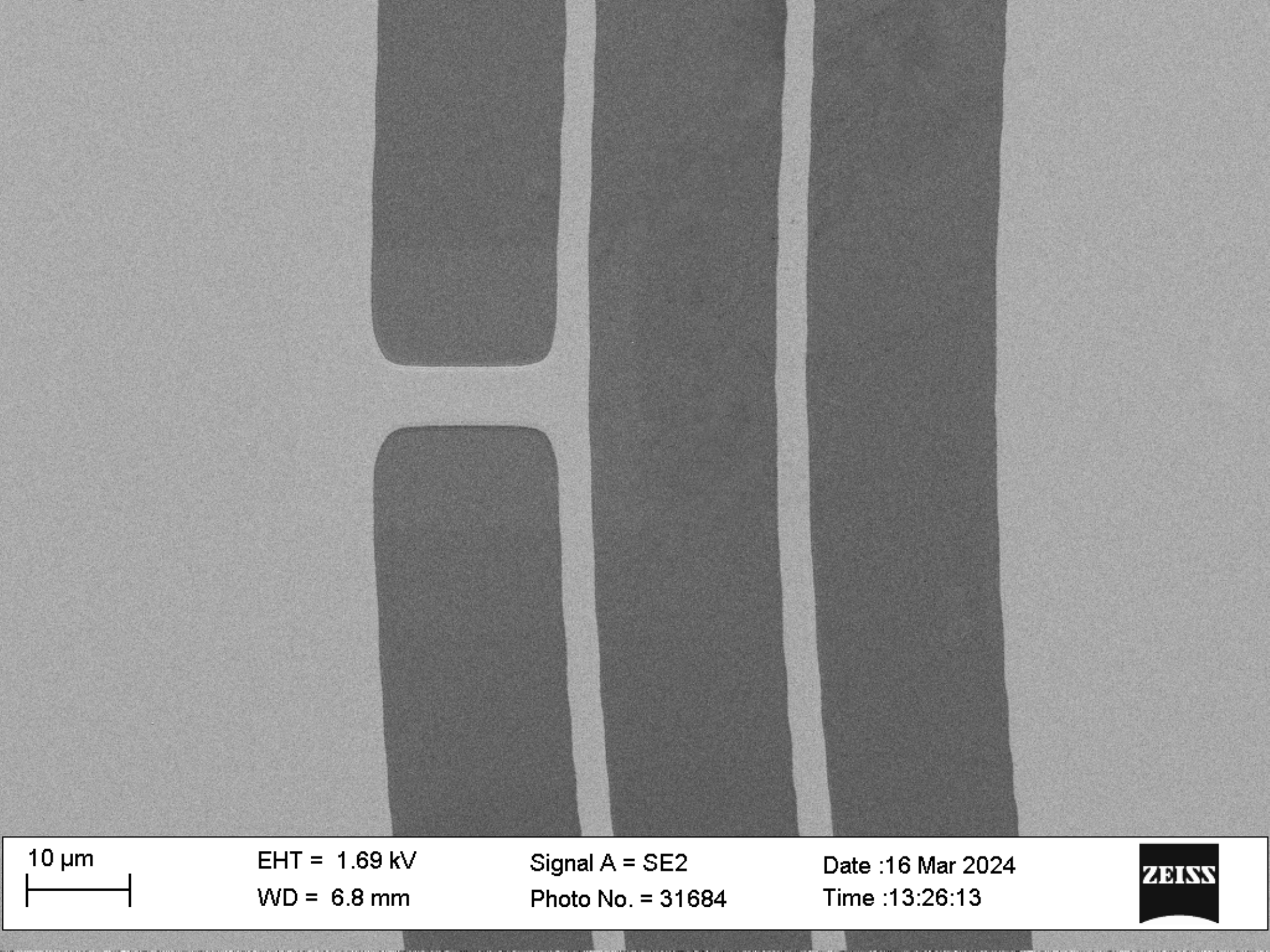}}
                \phantomsubcaption
        \label{Figure2b}
    \end{subfigure}
\caption{(a) A fabricated resonator under the microscope. (b) Zoom-in image showing the gap of the resonator obtained using a Scanning Electron Microscope (Zeiss Ultra SEM).}
\label{Figure2}
\end{figure}

The device described in this work is made of four different CSRRs on one chip, with identical ring dimensions other than their radii ranging from $r_{ext}=550 \,\mathrm{\mu m}$ to $612 \, \mathrm{\mu m}$. The design of the sample can be seen in Figure \ref{Figure1a}. It is made of a 9x5 mm sapphire substrate coated with a 42 nm (measured using Rigaku Smartlab XRD) NbTi film and etched to obtain the CSRR structure.\\
Split Ring Resonators (SRRs) \cite{Tan2019,Ricci_2006} and CSRRs \cite{McGregor2013} have been subject of interest in the past 20 years due to their good transmission and quality factor at microwave frequencies, within a small on-chip area. For a simple SRR the design can vary slightly, but is usually made of two concentric inductive loops of thickness $w$ and spacing $s$, with one or two gaps $g$ forming the parallel capacitor. The resonator design used in this work, which can be seen in Figure \ref{Figure1b}, is based on a superconducting CSRR structure \cite{Ricci_2006} of external radius $r_{ext}$ with one additional etched loop to make a continuous inner ring. The dimensions of $s$, $w$, $g$ and especially $r_{ext}$ can be changed to tune the quality factor and resonance frequency to the desired value and different resonator designs on the same chip can help obtain a comb-like amplification spectrum.
\noindent
The NbTi films have been deposited using a AJA Sputterer (ATC2200-V), have been measured to have a transition temperature $T_c\simeq$ 8.8 K \cite{Sweetnam2024} and a normal state resistivity $\rho_n\simeq$ 3.6$\cdot 10^{-7}$ $\mathrm{\Omega \cdot m}$. The pattern is then defined through optical photolithography (Microtech Laserwriter LW405B) and the sample is reactive ion etched (OIPT System 100 ICP RIE) in an SF\textrm{$_6$}+O\textrm{$_2$} environment to finalize the chip (see Figure \ref{Figure2a} and Figure \ref{Figure2b} for a zoom-in of the gap).\\
Using the values of $T_c$ and $\rho_n$ and calculating the geometric inductance of the CSRR \cite{McGregor2013}, the expected kinetic inductance fraction for a sample with $r_{ext}=612 \,\mathrm{\mu m}$, $w=5 \,\mathrm{\mu m}$, $s=5 \,\mathrm{\mu m}$ and $g=10 \,\mathrm{\mu m}$ (Resonator 1) has been found to be $\alpha_{exp}\simeq 57\,\%$. This value could be increased in future devices by making the films thinner or by changing the material used for the superconductor from NbTi to NbTiN or NbN, known to have a higher normal state resistivity but to be more difficult to fabricate reliably due to the need for nitrogen gas during sputtering \cite{Soldatenkova2021,Yang2018}. A sapphire substrate was chosen due to its desirable dielectric properties at these frequencies, combined with its relatively high thermal conductivity (compared to quartz).\\
The expected resonance frequencies of these resonators have been simulated using Ansys HFSS, falling within the range 34.5-38.5 GHz and spaced by about 1 GHz, with the largest radius giving 34.5 GHz. No significant hybridization of the resonances was observed, likely due to the large on-chip distance between the resonators. For designs with much larger numbers of CSRRs on the same chip, possible hybridization will have to be investigated in more detail. These simulations do not take into account any kinetic inductance and the measured resonance frequencies are expected to be shifted down in frequency by $f_0-f_m=(1-\sqrt{1-\alpha_{m}})f_0$, where $f_m$ is the measured resonance frequency (taking into account the kinetic inductance), $f_0$ is the un-shifted resonance frequency and $\alpha_{m}$ is the measured kinetic inductance fraction. By measuring $f_m$ and using the simulated $f_0$ a value for the kinetic inductance fraction has been derived: $\alpha_{m}\simeq 54\,\%$, in reasonably good agreement with the expected value of $57\,\%$.
The discrepancy between these two values could be due to several factors, including the geometric and film approximations used to calculate $\alpha_{exp}$, and limitations in the HFSS simulations. Films thinner than 42 nm have been explored but not used in this study, due to a lower yield and film uniformity. In order to improve the film quality, a lower sputtering rate could be used for future devices.

\subsection{Signal gain experimental setup}
\begin{figure}[tbp]
\centering
\includegraphics[width=0.75\linewidth]{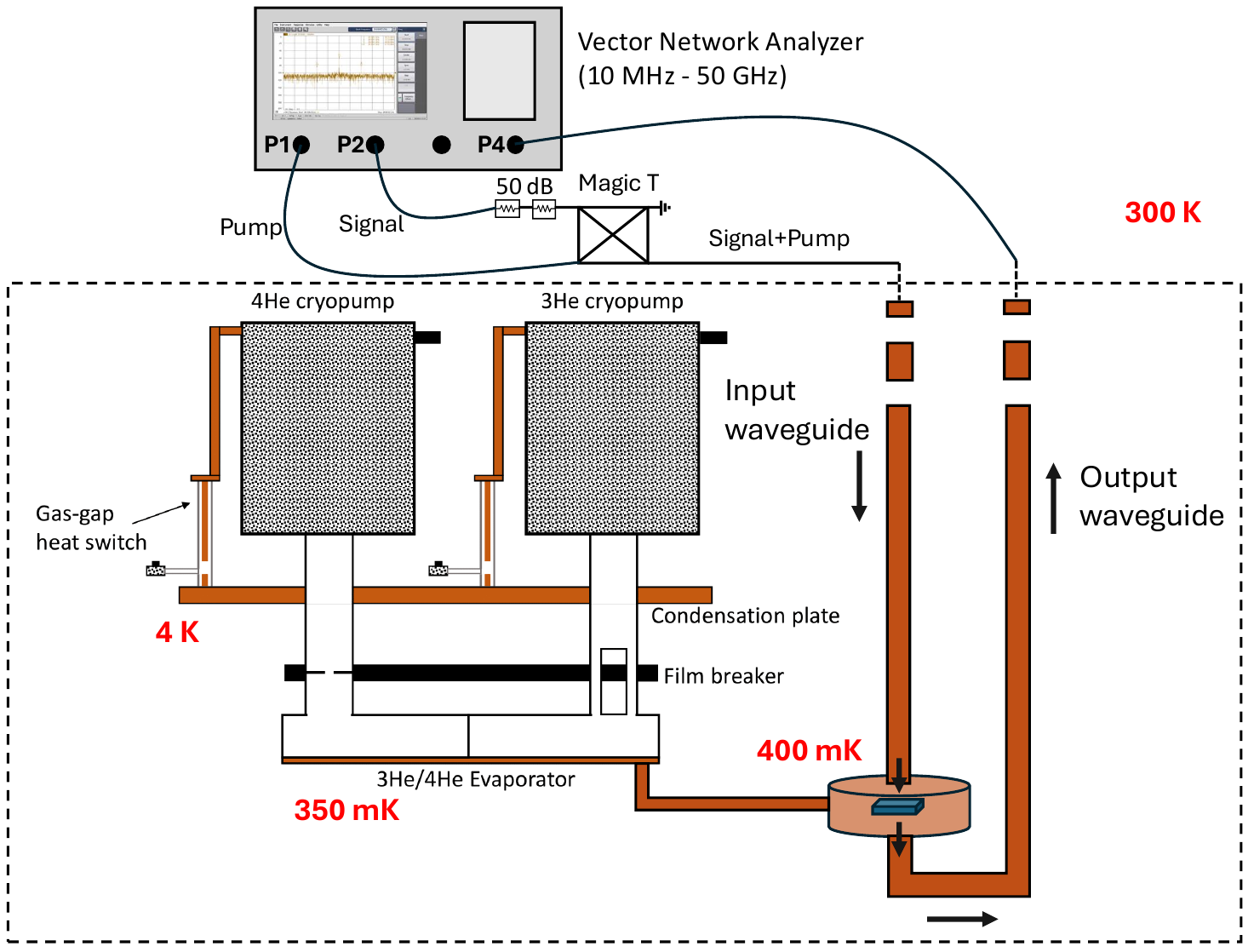} 
\caption{Diagram showing the gain test setup. Port 1 (P1) and Port 2 (P2) of the VNA provide the pump and signal tones that, after attenuation of the signal, are combined into the input waveguide chain using a Magic T coupler. Port 4 (P4) of the VNA reads the output tones which include the mixing products and the amplified signal. The sample is embedded in the sample holder and cooled down to 400 mK by the $^3$He - $^4$He sorption cooler.}
\label{Figure3}
\end{figure}

\begin{figure}[ht]
\centering
\includegraphics[width=0.4\linewidth]{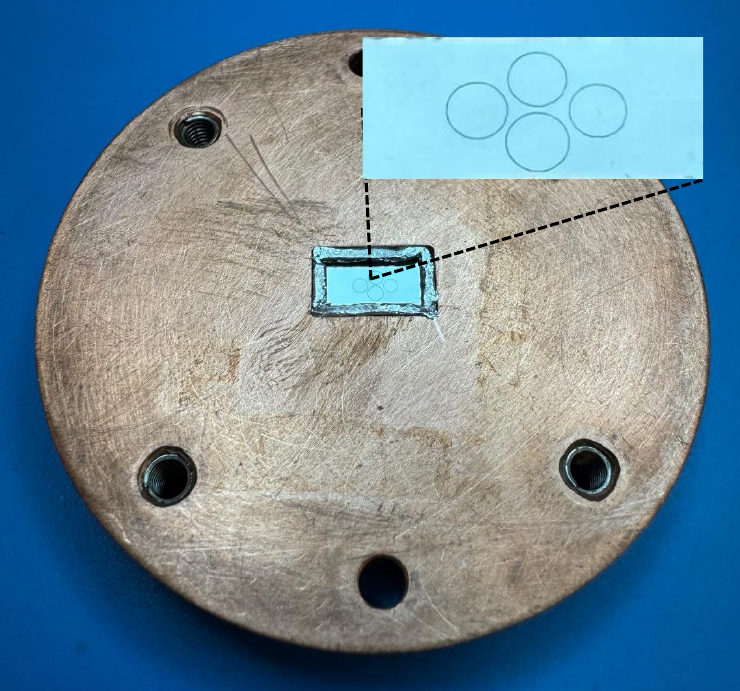}
\caption{The amplifying chip inside the sample holder.}
\label{Figure4}
\end{figure}

The signal gain has been measured using the setup shown in Figure \ref{Figure3}: two different ports (P1 and P2 in the figure) are used to provide the signal and pump tones. Using Resonance 3 in Figure \ref{Figure6} as an example, the pump is set at -30.2 dBm, which was identified to be close to the bifurcation power of this resonance, while the signal is set at -50 dBm and then further attenuated using a 30 dB external attenuator to -80 dBm. The pump and signal tones are then coupled using a Magic T into the input waveguide chain.
\noindent 
The amplifying chip is contained within the sample holder and cooled to 400 mK using a $^3$He - $^4$He single shot sorption cooler \cite{DallOglio1991}. After the mixing of the tones occurs the output signal is read out by Port 4 (P4) of the VNA. Figure \ref{Figure4} shows one of the fabricated chips inside the sample holder. Indium solder has been put at the edges of the slot of the sample holder to preserve the continuity of the fields and minimize the losses of the chain. The optical coupling achieved this way prevents any parasitic inductances that can arise due to wire bonding in classical planar SPAs.\\
Using this setup, the pump has been kept at resonance while the signal frequency has been swept across a 0.5 MHz band centred on the pump frequency. The mixing and gain of the amplifier are very sensitive to the pump power, with a deviation of 0.1 dB away from the ideal value of -30.2 dBm leading to 5-10 dB less signal gain.\\
The gain compression point of the amplifier has been studied using a similar setup to the one shown in Figure \ref{Figure3}. In this case the signal tone has been kept 10 kHz detuned from the pump tone, which was kept at resonance. The resonance used for this set of measurements was Resonance 2 in Figure \ref{Figure6}, with a center frequency of $\simeq$ 24.490 GHz. The pump power has been kept at bifurcation ($\sim$ -32 dBm) and the signal power swept from -60 to -110 dBm.

\subsection{Noise measurement setup}
\begin{figure}[ht]
\centering
\includegraphics[width=0.65\linewidth]{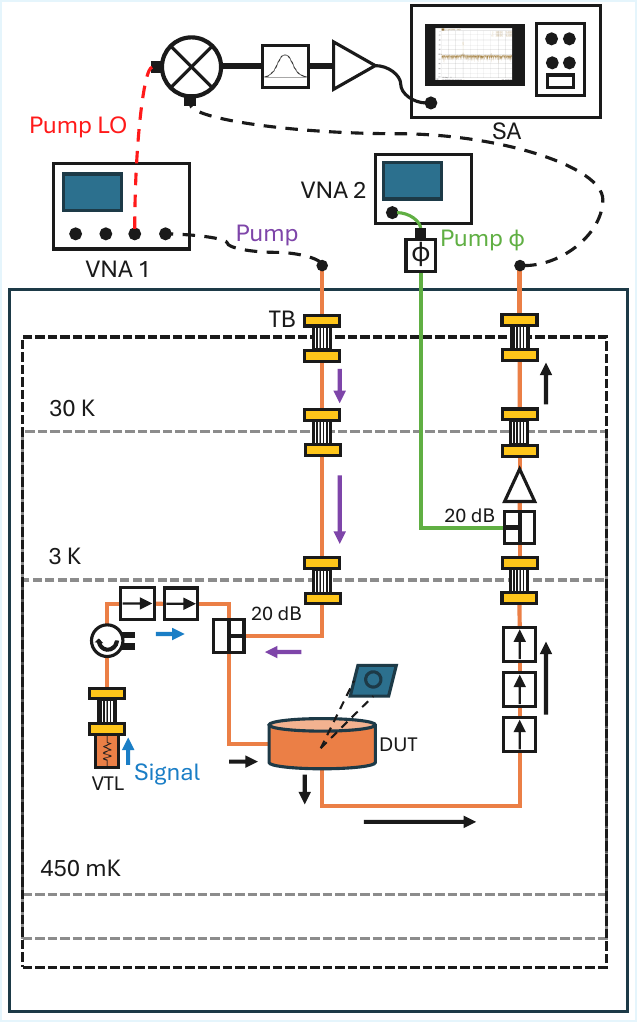}
\caption{Noise measurement setup. The pump (purple) is fed to the SPA using a 20 dB cryogenic directional coupler. A phase locked version of the same pump is used as the LO during the down-conversion in homodyne detection (red). A cryogenic stainless steel coaxial line is used to couple the -180 degrees phase-shifted pump before the HEMT amplifier (green).}
\label{Figure5}
\end{figure}

The experimental setup used to estimate the noise uses techniques that have become the state of the art of these types of measurements \cite{Tholen2014,Castellanos-Beltran2008}: homodyne detection and pump nulling. The setup is shown in Figure \ref{Figure5}. VNA 1 is used to provide the two phase-locked versions of the pump, one (purple) travels down the input waveguide chain and is fed to the Device Under Test (DUT) through a 20 dB directional coupler. The broadband noise is provided by a waveguide Variable Temperature Load (blue) coupled to the DUT by the other port of the directional coupler. A series of isolators and one double-junction circulator protect the Variable Temperature Load (VTL) from any reflections coming from the directional coupler or the DUT. These isolators are used for noise characterization only and they are not needed for correct operation of the device as a quantum amplifier.\\
The signal combined by the directional coupler (black) is then amplified by the DUT and travels towards the HEMT amplifier through the output section of the waveguide. Another directional coupler is used to combine the output noise (and the pump) with the 180 degrees phase shifted pump provided by VNA 2 (green). External phase shifters are used to adjust the phase of the pump used for nulling. Isolators protect the DUT from any signals reflected back by the coupler. After the amplification of the HEMT, the signal travels out of the waveguide chain and it is down-converted using a phase-locked version of the pump as the LO (homodyne detection). High pass filters are used to further suppress the pump tone and additional amplification is used at room temperature before feeding the signal to a spectrum analyser (Keysight Technologies MXA Signal Analyzer N9021B).
Homodyne detection is a technique that has been widely used in the past decade to measure the noise of SPAs \cite{Castellanos-Beltran2008}\cite{Tholen2014}. Using a copy of the pump as the Local Oscillator (LO) of a mixer in down-conversion, the entire signal can be shifted to or near DC. This facilitates filtering and gives the possibility of trying to achieve noise squeezing in the right conditions. Squeezing was not explored in this work due to the limitations of the test setup.\\
Pump nulling is another tool widely used for the testing of four-wave mixing devices, often used to avoid compression of subsequent amplifiers due to the strong pump. In our case, the pump nulling circuit was tested at room temperature and then implemented in the final setup. Significant nulling was achieved during these measurements, with between 80-90 dB of stable nulling of the pump. Above these values the nulling became unstable, likely due to minor phase or power fluctuations.\\
Each temperature stage of waveguide components was kept isolated from the rest using waveguide Thermal Breaks \cite{Melhuish2016} (TB in Figure \ref{Figure5}). The waveguide chains were made of copper and very well anchored to each temperature stage. Their great thermal properties ensure an ideal low-noise electromagnetic environment of the input chain feeding the pump to the DUT. \\
To set up for noise measurements, with the fridge cold, the SPA was driven close to bifurcation with a pump tone at resonance of -12.4 dBm (-32.4 dBm after the directional coupler) and a test tone was used to measure the signal gain ($\simeq$ 20 dB). Then the pump nulling circuit was tuned to achieve at least 60 dB of stable pump attenuation. The frequency of the LO pump tone was chosen to be about 30 kHz smaller than the pump fed to the DUT, in near-homodyne detection. To detect the noise, a narrow bandwidth of 3 kHz was chosen centred 100 kHz away from the pump tone and integrated over this bandwidth.\\
After this set up, the temperature of the VTL was varied slowly using a weak heater (1 M$\mathrm{\Omega}$) and waiting several minutes at each temperature step. A cryogenic Heat Switch was used whenever rapid cooling of the VTL was necessary, while a weak thermal link to the fridge was used to help with thermalization. The noise was integrated and averaged at least 100 times for each data point to capture the small changes in output power. Each uncertainty was chosen as the average fluctuation of this value over several minutes. During this process, the temperature of the DUT was also monitored and it showed minimal fluctuations of the order of 10 mK.\\
It is important to note that the noise temperature of the VTL measured this way is not the actual equivalent noise power at the input of the DUT due to the attenuation of the waveguide chain, isolators and circulator connecting the VTL to the DUT (see Figure \ref{Figure5}). For this reason, two subsequent calibration runs have been conducted to extract both the attenuation of the input chain (providing the pump tone) and of the input line of the VTL. The first was measured to be $\sim$ -3.543 dB, while the second $(-0.764\pm0.002)$ dB, which was subsequently used to correct the noise temperature on the x-axis in Figure \ref{Figure9}.

\section{Results and discussions}\label{Results and discussions}
\subsection{S-parameters}
\begin{figure}[ht]
\centering
\includegraphics[width=0.65\linewidth]{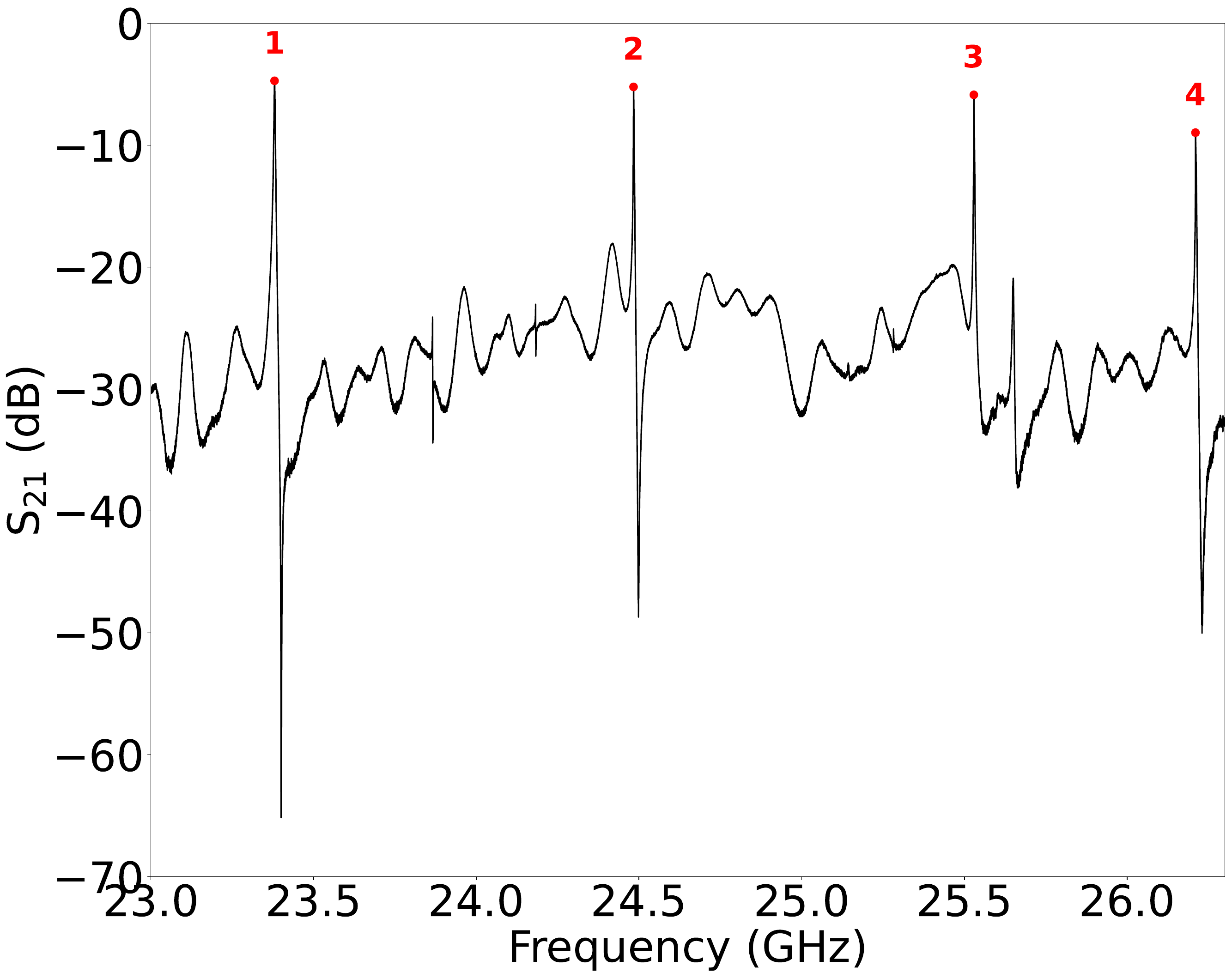} 
\caption{Transmission measured at 400 mK from the CSRR chip. All the four resonances are visible, with a transmission ranging from -8 to -5 dB and a quality factor of about $1\cdot10^4$.}
\label{Figure6}
\end{figure}

The transmission has been measured at 400 mK using a Vector Network Analyser (VNA, Keysight PNAx N5245B) and can be seen in Figure \ref{Figure6}. All four resonances are visible in the figure, showing high transmission ranging from -8 to -5 dB (with -4.5 dB being the measured attenuation of the input-output chain) and a quality factor of $7\cdot10^3$ for the first two resonances and $1\cdot10^4$ for the last two. As per design, the resonances are spaced by roughly 1 GHz. Some small spurious resonances can also be observed, which are assumed to be related to partial structures of the CSRRs, like a single-ring resonance.
\noindent

\subsection{Gain performance}
\begin{figure}[ht]
\centering
\includegraphics[width=0.65\linewidth]{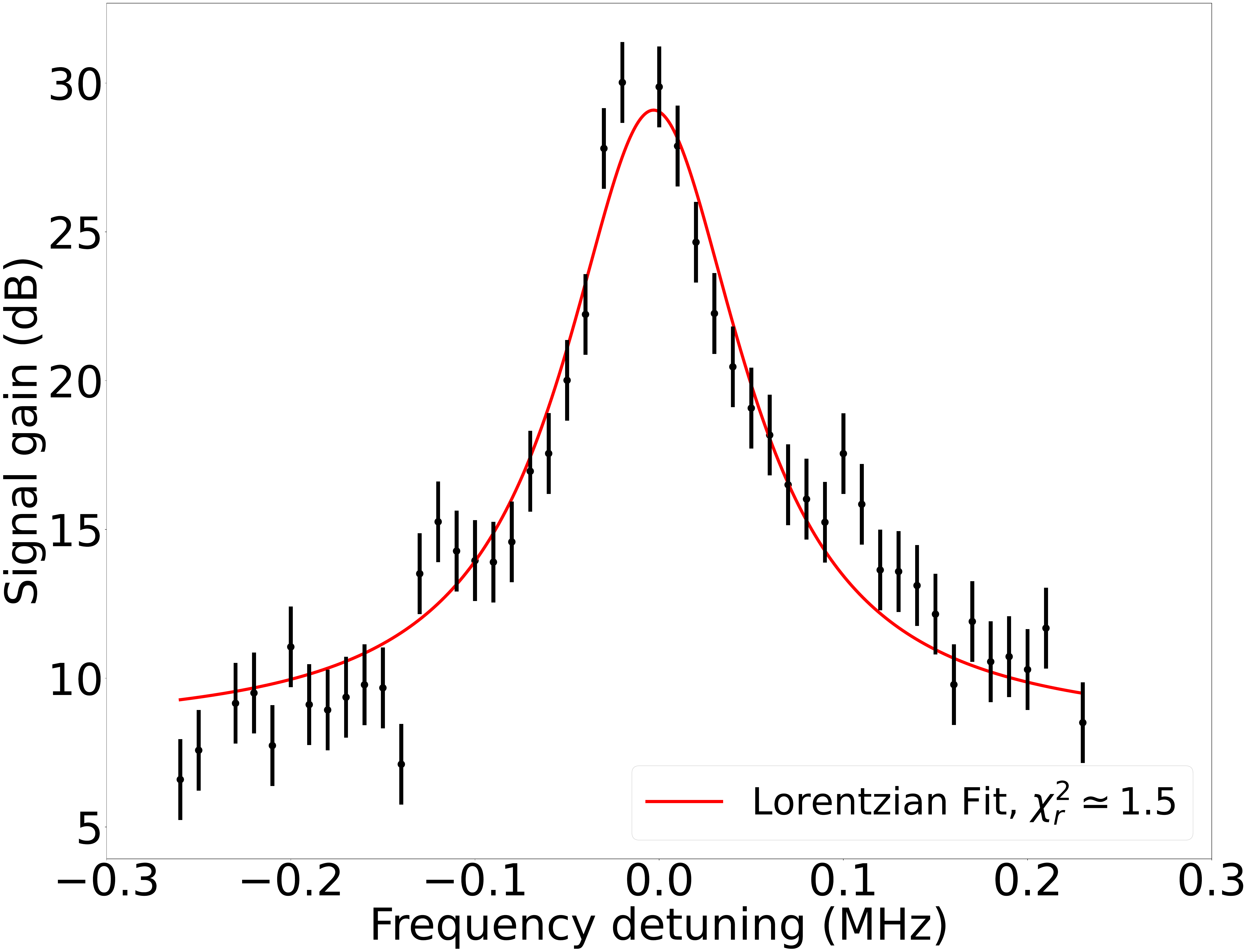}
\caption{Signal gain measured at 400 mK fixing the pump frequency at resonance and sweeping the signal across a 0.5 MHz band centred at 25.529 GHz (the center of the third resonance in Figure \ref{Figure6}). A Lorentzian fit has been performed on the data with a reduced Chi-square of 1.5, suggesting a reasonable match to the data.}
\label{Figure7}
\end{figure}

\begin{figure}[ht]
\centering
\includegraphics[width=0.65\linewidth]{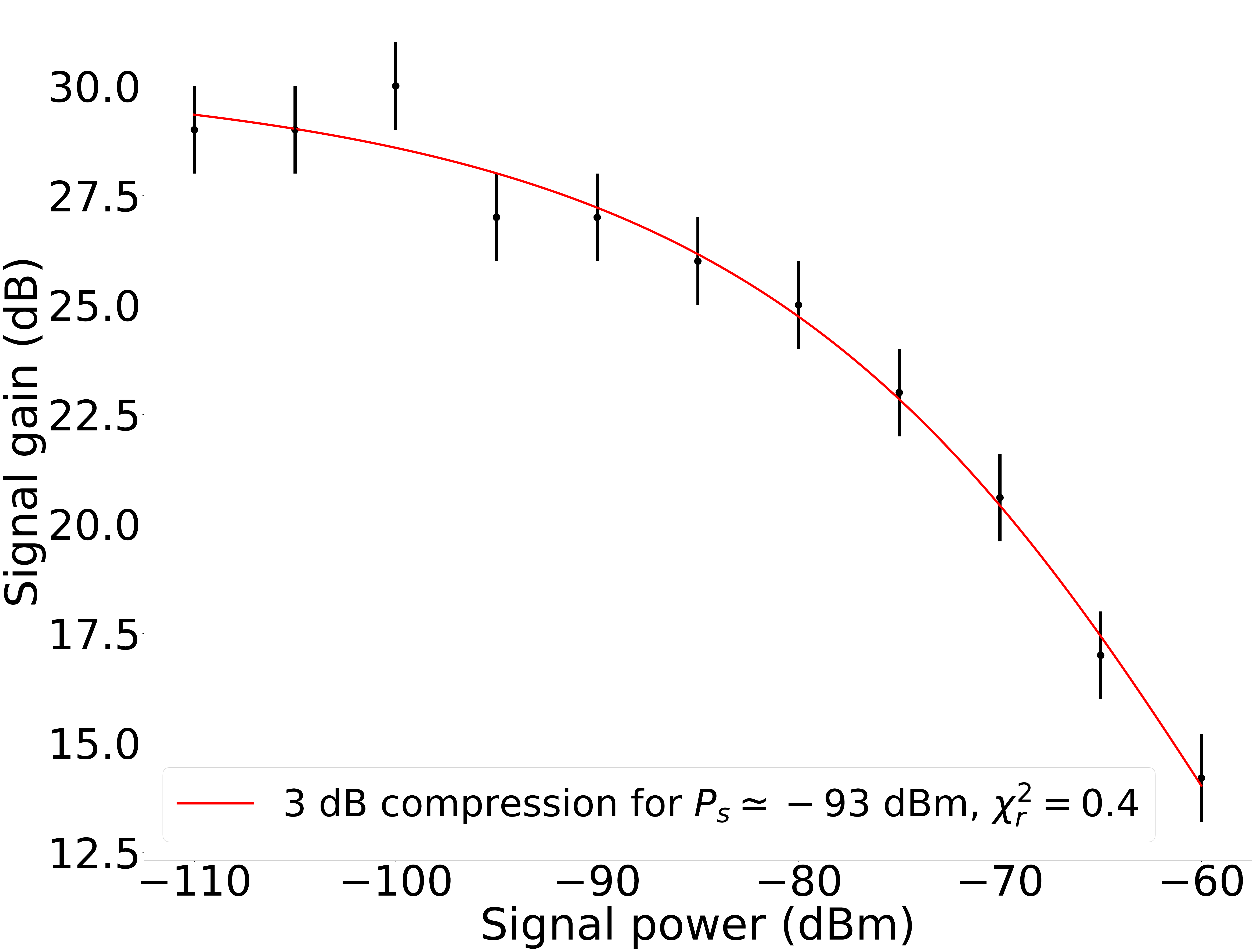}
\caption{Gain compression plot for Resonance 2 in Figure \ref{Figure6}. The 3 dB compression point is achieved for an input signal power of -93 dBm and the gain saturates at about 30 dB. The reduced Chi-square of 0.4 shows that the uncertainties are slightly overestimated.}
\label{Figure8}
\end{figure}

Using the measurement setup described in Figure \ref{Figure3}, the gain performance of each one of the four resonators has been measured at 400 mK and is shown for Resonance 3 in Figure \ref{Figure7}.
A maximum signal gain of 30 dB is measured, with a 3 dB bandwidth of $\sim$ 150 kHz, leading to a gain-bandwidth product of $\sim$ 150 MHz. Subtracting the loss of the resonator at the peak (-1 dB) from the signal gain leads to 29 dB of insertion gain. All of the resonances shown in Figure \ref{Figure6} exhibit a strong non-linearity with greater than 25 dB of maximum signal gain. The uncertainty in the gain in Figure \ref{Figure7} and Figure \ref{Figure8} has been estimated by repeating the sweep several times.
\noindent
The 3 dB gain compression point has been measured for Resonance 2 (Figure \ref{Figure8}) and identified to occur at -93 dBm, with the gain saturating at 30 dB, as expected due to the similarity with the resonance in Figure \ref{Figure7}. The reduced Chi-square obtained for this fit suggests overestimated uncertainties. These have been derived as the observed signal gain fluctuations over time at each signal power. These fluctuations can depend on various factors, like a slight change in the operating point, signal instability in highly non-linear regimes and resonance instability. To take all of these systematics into account, they have been estimated conservatively. The compression point of the other resonances has not been measured but a similar behaviour is expected due to the similar design and gain performance.

\subsection{Noise measurements}
\begin{figure}[ht]
\centering
\includegraphics[width=0.65\linewidth]{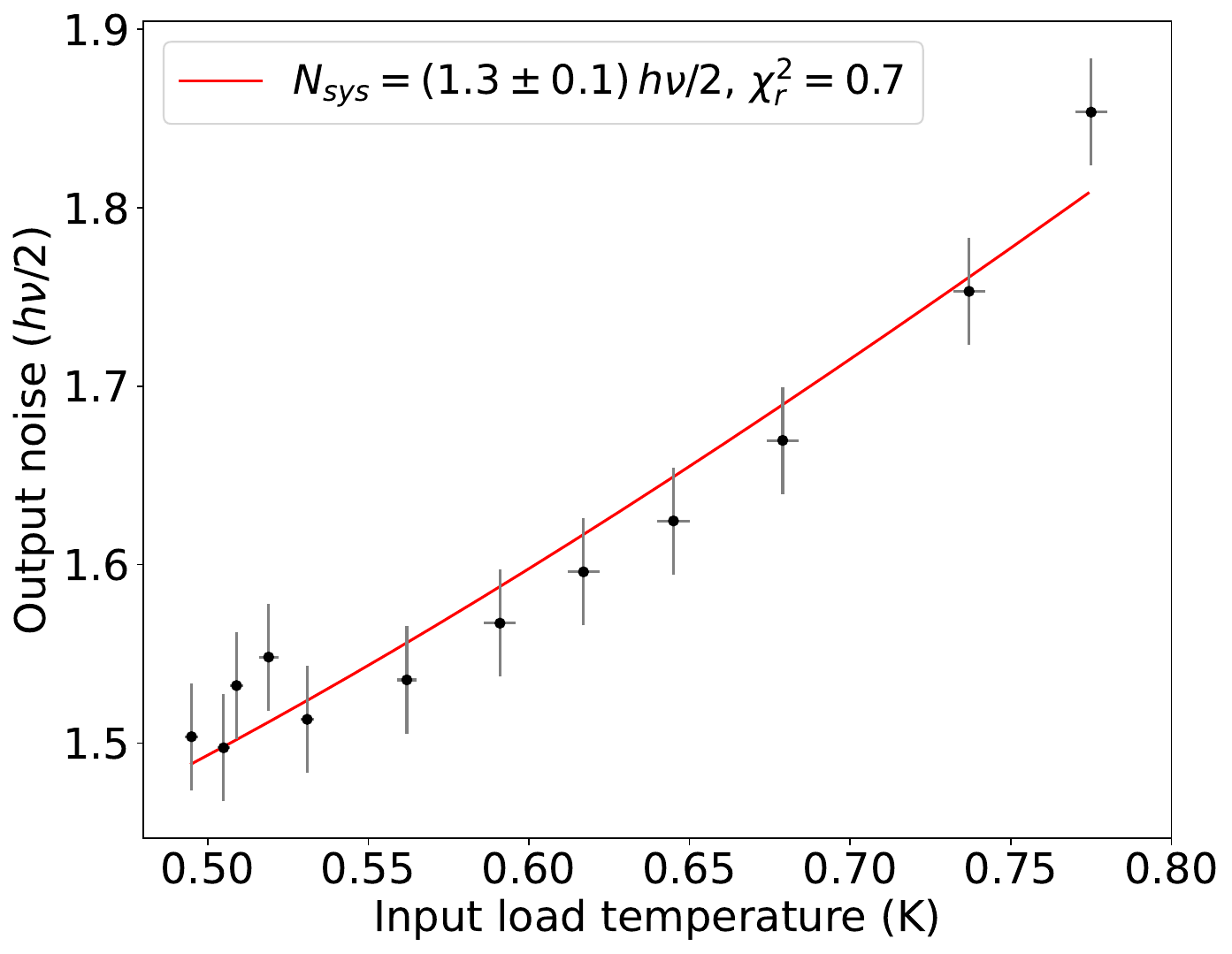}
\caption{Output noise power as a function of the noise temperature in input at the DUT. The estimated system noise is $(1.3 \pm 0.1)\:h \nu/2$ and the reduced Chi-square is 0.7.}
\label{Figure9}
\end{figure}

The noise power coming from Resonance 4 measured as a function of the temperature of the VTL is shown in Figure \ref{Figure9} in units of number of half quanta ($h\nu/2$).
\noindent
The function used to fit the data points is \cite{Tholen2014}:

\begin{equation}
N_{TOT}(T)= N_{ADD} +\mathrm{coth}\left[\frac{h \nu_0}{2k_BT}\right]
\label{Equation3}
\end{equation}
where $T$ is the noise temperature at the input of the SPA, $\nu_{0}=26.1737$ GHz is the resonance frequency and $N_{ADD}$ the added system noise above vacuum. The extrapolated plateau obtained from the fit in Figure \ref{Figure9} gives a total system noise of $(1.3 \pm 0.1)\:h \nu/2$, very close to the quantum limit. This total noise includes the noise of the HEMT amplifier suppressed by the gain of the SPA, which was measured to be around 20 dB during these measurements. Driving the amplifier to its maximum gain performance is normally considered undesirable for noise measurements, due to instability during data acquisition. \\
The HEMT amplifier used for these measurements (a MMIC from JPL Pasadena) was measured to provide 30 dB of gain with a noise power (at an operating temperature of 5 K) \cite{McCulloch2017}: $N_{HEMT} \simeq 14.3 \: h  \nu/2$. This noise suppressed by the gain of the SPA leads to $N_{HEMT}/G(L) \simeq 0.143 \: h  \nu/2$ (where $G(L)$ is the gain in linear units) accounting for about half the extrapolated noise above the quantum limit and suggesting a noise added by the SPA: $N_{SPA}=(1.2 \pm 0.1)\,h\nu/2$, corresponding to an equivalent noise temperature of 0.75 K. The section of waveguide connecting the VTL to the DUT was kept at the same temperature as the VTL's starting point and would have contributed minimally to the overall noise. Additional cryogenic and warm components also contribute to the measured noise but they are suppressed by both the gain of the SPA and the HEMT, rendering them negligible. The additional noise above the quantum limit could be due to several factors, including: calibration uncertainty, impedance mismatch, residual pump noise and thermal occupation. The noise performance of the other resonators was not investigated in this manuscript, but it is expected to be qualitatively similar, due to their identical design (CSRR) and comparable gain performance. The near quantum limited noise performance of the device is consistent with the majority of the non-linearity being of reactive nature, due to the current-dependent kinetic inductance described in Equation \ref{Equation1}.\\
Table \ref{Table1} shows a comparison between the noise performance of the design presented in this work and other SPA designs. One in particular (Ref \cite{Hao2026}) reports an amplifier added noise of $\sim$ 2 photons at similar frequencies.
\begin{table}[h]
\begin{ruledtabular}
\begin{tabular}{lcccc}
Reference & Type & Frequency (GHz) & Noise ($h\nu/2$)\\
\hline
Ref. \cite{Macklin2015}& JTWPA & 4--10 & $ 1.5\pm0.1$  \\
Ref. \cite{Kaufman2025} & RFSQUID--JPA & $\sim 5.7^*$ & $\sim 1.8$  \\
Ref.\cite{Hao2026} & JPA & 21--23$^*$ & $\sim 4$  \\
This work & CSRR SPA & 23--26$^*$ & $1.2 \pm 0.1$ \\
\end{tabular}
\end{ruledtabular}
\caption{Comparison of the noise performance of superconducting parametric amplifiers at different frequencies. An asterisk ($^*$) has been used to denote narrow-band amplifiers. }
\label{Table1}
\end{table}

\section{Conclusions}
This work has demonstrated quantum limited microwave amplification without the need for a dilution refrigerator working in the 10-20 mK temperature range.\\
The device presented, made of four K band CSRRs optically coupled and embedded in a waveguide, has shown a maximum signal gain of 30 dB for all resonators with correspondingly high insertion gain (29 dB) due to their high transmission (-1 to -5 dB). The signal gain as a function of the input signal power at bifurcation has provided an estimated 3 dB compression point of -93 dBm.\\
The noise power of the fourth resonator (in frequency) has been measured at 400 mK, giving an estimated amplifier noise of $(1.2 \pm 0.1)\:h \nu/2$, which is very close to the quantum limit. In addition, the optical coupling of these resonators gives the opportunity of scaling up the number of resonators on one chip, achieving a comb-like quantum limited amplifying chip.\\
These results show that, working at higher frequencies allows the use of simpler cryogenics to access the quantum regime, which could have a significant impact on many fields that use these technologies, from astrophysics detection to quantum computing. Future iterations of these devices will have to be developed with the intent of exploring even higher frequencies and to include a method for frequency tuning, both crucial for future Axion dark matter experiments and quantum technologies.

\section*{Funding}
This work was supported by the Science and Technology Facility Council (STFC, ST/X001229/1).

\section*{Acknowledgements}
Device fabrication was performed at the National Graphene Institute (NGI) at the University of Manchester.

\section*{Author contributions}
V. G. conceived and designed the work presented, worked on the theoretical model used to guide the fabrication, performed all of the measurements, analysed the data, made the figures and wrote the manuscript. T. S. studied earlier prototypes, contributing to the design, measurements and data analysis, laying the ground work that supported this publication. B. M. fabricated the devices and optimised the processing. M. A. McC. created the initial designs, recognised the potential of these devices and proposed their use in this application. L. P. supervised the project as the team leader.

\section*{Competing interests}
The authors declare no competing interests.

\section*{Data availability}
The datasets generated during this study are available from the corresponding author upon reasonable request.

\section*{Code availability}
No custom code was used in this study.

\bibliography{References}

\appendix

\section{Noise measurements data}
The calibrated and normalized data used for Figure \ref{Figure9} is shown in Table \ref{Table1}, with the output noise in units of half quanta and the input load temperature in Kelvin.

\begin{table}[h]
\begin{tabular}{lcccc}
$T_{\mathrm{load}}$ (K) & $N_{\mathrm{out}}$ ($h\nu/2$) & $\delta T$ (K) & $\delta N_{\mathrm{out}}$ \\
\hline
0.495 & 1.504 & 0.002 & 0.030 \\
0.505 & 1.497 & 0.002 & 0.030 \\
0.509 & 1.532 & 0.002 & 0.030 \\
0.519 & 1.548 & 0.003 & 0.030 \\
0.531 & 1.513 & 0.002 & 0.030 \\
0.562 & 1.535 & 0.003 & 0.030 \\
0.591 & 1.567 & 0.005 & 0.030 \\
0.617 & 1.596 & 0.005 & 0.030 \\
0.645 & 1.624 & 0.005 & 0.030 \\
0.679 & 1.670 & 0.005 & 0.030 \\
0.737 & 1.753 & 0.005 & 0.030 \\
0.775 & 1.854 & 0.005 & 0.030 \\
\end{tabular}
\caption{Output noise as a function of the input load temperature.}
\label{Table1}
\end{table}
\end{document}